\documentclass[twocolumn,superscriptaddress,floatfix,preprintnumbers,prl ,nofootinbib]{revtex4}
\usepackage[colorlinks=true,breaklinks=true]{hyperref}
\usepackage[utf8]{inputenc}
\hypersetup{allcolors=[rgb]{0.0 0.0 0.6},linkcolor=[rgb]{0.75 0.05 0.05}}
\usepackage{mathtools}
\usepackage{float}
\usepackage{siunitx}
\usepackage{letltxmacro}
\LetLtxMacro{\oldcite}{\cite}
\renewcommand{\cite}[1]{\mbox{\oldcite{#1}}}

\DeclareSIUnit\electronvolt{e\kern-.05em V}
\DeclareSIUnit\parsec{\text{pc}}
\DeclareSIUnit\clight{\text{\ensuremath{c}}}

\usepackage{orcidlink}

\long\def\exclude#1{}

\def\lya{Lyman-$\alpha$\,}

\newcommand{\beq}{\begin{equation}}
\newcommand{\eeq}{\end{equation}}

\def\ga{\,\,\raise0.14em\hbox{$>$}\kern-0.76em\lower0.28em\hbox
{$\sim$}\,\,}

\newcommand{\peak} {$k_{\star}/k_{\rm peak}$\,} 

\allowdisplaybreaks

\bibpunct{[}{]}{,}{n}{}{,}

\usepackage[colorinlistoftodos]{todonotes}

\begin{document}

\title{Constraints on Primordial Magnetic Fields from the Lyman-$\alpha$ forest}

\author{Mak Pavi\v{c}evi\'{c}\,\orcidlink{0000-0002-0570-4256}}
\email{mak.pavicevic@sissa.it}
\affiliation{SISSA - International School for Advanced Studies, Via Bonomea 265, I-34136 Trieste, Italy}
\affiliation{IFPU, Institute for Fundamental Physics of the Universe, Via Beirut 2, I-34151 Trieste, Italy}
\affiliation{INFN, Sezione di Trieste, Via Valerio 2, I-34127 Trieste, Italy}
\author{Vid Ir\v{s}i\v{c}\,\orcidlink{0000-0002-5445-461X}} \email{v.irsic@herts.ac.uk}
\affiliation{SISSA - International School for Advanced Studies, Via Bonomea 265, I-34136 Trieste, Italy}
\affiliation{IFPU, Institute for Fundamental Physics of the Universe, Via Beirut 2, I-34151 Trieste, Italy}
\affiliation{INFN, Sezione di Trieste, Via Valerio 2, I-34127 Trieste, Italy}
\affiliation{INAF - Osservatorio Astronomico di Trieste, Via G. B. Tiepolo 11, I-34143 Trieste, Italy}
\affiliation{KICC - Kavli Institute for Cosmology Cambridge, Madingley Road, CB3 0HA Cambridge, United Kingdom}
\affiliation{Center for Astrophysics Research, Department of Physics, Astronomy and Mathematics, University of Hertfordshire, College Lane, Hatfield AL10 9AB, UK}
\author{Matteo Viel\,\orcidlink{0000-0002-2642-5707}}
\email{viel@sissa.it}
\affiliation{SISSA - International School for Advanced Studies, Via Bonomea 265, I-34136 Trieste, Italy}
\affiliation{IFPU, Institute for Fundamental Physics of the Universe, Via Beirut 2, I-34151 Trieste, Italy}
\affiliation{INFN, Sezione di Trieste, Via Valerio 2, I-34127 Trieste, Italy}
\affiliation{INAF - Osservatorio Astronomico di Trieste, Via G. B. Tiepolo 11, I-34143 Trieste, Italy}
\affiliation{ICSC - Centro Nazionale di Ricerca in High Performance Computing,
Big Data e Quantum Computing,
Via Magnanelli 2, Bologna, Italy}

\author{James S. Bolton\,\orcidlink{0000-0003-2764-8248}}
\affiliation{School of Physics and Astronomy, University of Nottingham, University Park, Nottingham, NG7 2RD, UK}

\author{Martin G. Haehnelt\,\orcidlink{0000-0001-8443-2393}}
\affiliation{Institute of Astronomy and Kavli Institute for Cosmology, University of Cambridge, Madingley Road, Cambridge CB3 0HA, UK}

\author{Sergio Martin-Alvarez\,\orcidlink{0000-0002-4059-9850}}
\affiliation{Kavli Institute for Particle Astrophysics \& Cosmology (KIPAC), Stanford University, Stanford, CA 94305, USA}

\author{Ewald Puchwein\,\orcidlink{0000-0001-8778-7587}}
\affiliation{Leibniz-Institut für Astrophysik Potsdam (AIP), An der Sternwarte 16,
14482 Potsdam, Germany}

\author{Pranjal Ralegankar\,\orcidlink{0000-0001-7696-5878}}
\affiliation{SISSA - International School for Advanced Studies, Via Bonomea 265, I-34136 Trieste, Italy}
\affiliation{IFPU, Institute for Fundamental Physics of the Universe, Via Beirut 2, I-34151 Trieste, Italy}

\begin{abstract}
We present the first constraints on primordial magnetic fields from the Lyman-$\alpha$ forest using full cosmological hydrodynamic simulations. At the scales and redshifts probed by the data, the flux power spectrum is extremely sensitive to the extra power induced by primordial magnetic fields in the linear matter power spectrum, at a scale that we parametrize with $k_{\rm peak}$. We rely on a set of more than a quarter million flux models obtained by varying thermal, reionization histories and cosmological parameters. We find a hint of extra power that is well fitted by the PMF model with $B\sim 0.2$ nG, corresponding to $k_{\rm peak}\sim 20$ Mpc$^{-1}$. However, when applying very conservative assumptions on the modelling of the noise, we obtain a 3$\sigma$ C.L. lower limit $k_{\rm peak}> 30$ Mpc$^{-1}$ which translates into the tightest bounds on the strength of primordial intergalactic magnetic fields: $B < 0.30$ nG (for a fixed, nearly scale-invariant $n_{\rm B}=-2.9$).
\end{abstract}

\maketitle

\textit{Introduction.}--- Primordial magnetic fields (PMFs) could provide invaluable insights on new fundamental physics in the early Universe like inflation or phase transitions \cite{Subramanian:2015lua,Vachaspati:2020blt}. Their impact in early structure formation takes place primarily at small cosmological scales: the Lorentz force acting on baryon plasma induces growth of perturbations, an effect which has been investigated at a linear level \cite{zeldovich70,Kim:1994zh,Shaw:2010ea} or in the mildly non-linear regime \cite{Mtchedlidze:2021bfy,ralegankar24b}. Magnetic fields are observationally constrained, even at cosmological distances, using a variety of probes including gamma-ray observations of Blazars \cite{Finke:2015ona,HESS:2014kkl,VERITAS:2017gkr}, the Cosmic Microwave Background  (CMB) \cite{paoletti11,Jedamzik:2013gua,Planck:2015zrl,jedamzik19}, and, more in the context of structure formation, reionization \cite{pandeysethi:2015,Katz:2021,paoletti22,jedamzik25}, dwarf galaxies \cite{Sanati_2020,sanati24}, line intensity mapping \cite{adi23}, cosmic voids \cite{neronov10}, Faraday rotation measurements \cite{neronov24} and observations of high$-z$ galaxies \cite{ralegankar24}. Many of these are affected by large uncertainties in the astrophysical modelling.
Recently, there have been several efforts in the context of developing physics rich hydrodynamic simulations, which might include magneto-hydrodynamics, to simulate {\it ab initio} the development of magnetic fields from astrophysical sources and to trace their physical properties in a variety of environments through cosmic epochs \cite{dolag05,martin-alvarez20,Vazza:2017qge,Vazza:2017mbz,Katz:2021,kannan22}. These attempts allow to reach very small scales making it possible to provide new constraints on PMFs. 

Among the different probes, a unique key-role is played by the Lyman-$\alpha$ forest, the main observable manifestation of the intergalactic medium (IGM), produced by intervening neutral hydrogen along the line-of-sight to distant quasars. The forest has been shown to provide tight constraints on dark matter free-streaming, neutrino masses, cosmological parameters, primordial black holes, including, at large scales, geometrical constraints via baryonic acoustic oscillations measurements  and dark energy \cite{irsic24,neutrinolya,slosar13,murgia19,hooper22,villa23,goldstein}. New physics has been mainly searched in the context of models which provide a suppression of power at small scales, here, instead, we are interested in quantifying how an increase of power in the initial conditions will propagate to our final observable.

In this {\it Letter}, we argue that the Lyman-$\alpha$ forest could be the ideal way to constrain cosmological PMFs for two reasons: firstly, it explores a vast variety of scales, and is distinctively capable of reaching very small scales by using the one-dimensional (1D) flux power spectrum; secondly, it probes the filamentary cosmic web in environments at around the mean density, far from galaxies, where the impact of magnetic fields generated by astrophysical sources should be minimal \cite{katz19,martin-alvarez21}.
Compared to previous works, in which the Lyman-$\alpha$ forest has been used to provide quantitative constraints on PMFs \cite{shawlewis12,chong13,kahnias13,Pandey2013}, we perform a full extensive set of hydrodynamic simulations that follow the evolution of the perturbations at very non-linear scales. These simulations should provide a very accurate state-of-the-art description of the physical properties of the diffuse cosmic web and the main summary statistics that we will use: the 1D transmitted flux power.
\textit{PMFs and the matter power spectrum.}--- 
We consider stochastic configurations of a non-helical primordial magnetic field $\mathbf{B}(x)$, such that its components $B_i(x)$ have a Gaussian distribution. All the units for the magnetic field strength and scales or wavenumbers are comoving units. The magnetic power spectrum $P_B(k)$ 
is assumed to be in a form of a power-law, i.e., $P_{\rm B} (k) \propto B^2_{\rm 1Mpc} k^{n_B}\exp(-k^2\lambda_D^2)$, where $n_{\rm B}$ is the spectral index, $B_{\rm 1Mpc}$ is the amplitude of the magnetic field smoothed over 1 Mpc, 
and the exponential factor ${\rm exp}(-k^2 \lambda_{\rm D}^2)$ phenomenologically encapsulates the suppression of power due to turbulent damping.\footnote{In reality, turbulence causes a power-law damping of the magnetic fields \cite{ralegankar24b, Banerjee:2004df}. As the exact shape of damping is largely unimportant for evaluating the matter power spectrum, we follow the convention in the literature and assume exponential damping for simplicity.} The damping scale is determined by \cite{Jedamzik:1996wp,Subramanian:1997gi,ralegankar24},
\begin{equation}
    \lambda_{\rm D} \sim \left[0.1\, \left( \frac{B_{\rm 1\,Mpc}}{\rm 1\, nG}\right) \right]^{2/(n_B +5)} \,\, {\rm Mpc} \,.
    \label{eq:lambdaD}
\end{equation}

In the presence of magnetic fields, divergence of the Lorentz force, $S(x) \propto \nabla \cdot [\mathbf{B}(x) \times \nabla \times \mathbf{B}(x)]$, directly sources perturbations in the baryon fluid \cite{wasserman78}, which then gravitationally sources dark matter perturbations. The PMF-induced matter power spectrum is of the form \cite{Kim:1994zh,Subramanian:2015lua}
\begin{equation}
    \Delta_{\rm m, \,PMF}^2 \propto 
    \left( \frac{B_{\rm 1\,Mpc}}{\rm 1\, nG} \right)^4   \left( \frac{k}{\rm 1\, Mpc ^{-1}} \right)^{2n_B +10} \, e^{-2k^2 \lambda^2_{\rm D}} \, .
\end{equation}
The total matter power spectrum is given by the sum P$_{\rm PMF}$+ P$_{\rm \Lambda CDM}$, assuming that inflationary and PMF-induced perturbations are uncorrelated.
The wave number at which the $\Delta^2_{\rm m}(k)$ reaches its maximum is
\begin{equation}
k_{\rm peak}=\left[0.1\, \kappa(n_B) \left( \frac{B_{\rm 1\,Mpc}}{\rm 1\, nG}\right) \right]^{\frac{-2}{(n_B +5)}} \sqrt{\frac{n_{\rm B}+5}{2}} \,\,\, {\rm Mpc^{-1}}\,,
\label{eqkpeak}
\end{equation}
where $\kappa(n_B)$ is an $\mathcal{O}(1)$ number that precisely specifies $\lambda_D$ for given $B_{\rm 1\, Mpc}$ and $n_B$. The exact value of $\kappa$ has been a subject of recent debate \cite{ralegankar24b}. In this study, we use the matter power spectrum and $\kappa$ derived in \cite{ralegankar24}, where $\kappa(-2.9)=0.88$ and $\kappa(-2.0)=1.79$.\footnote{Ref.~\cite{ralegankar24} only provides the expression for the baryon power spectrum in Eq.~2.23. We obtain matter power spectrum using
\begin{equation}
    \Delta^2_{\rm m, PMF}(k,z)=\left[\frac{\Omega_{\rm b}}{\Omega_{\rm m}}+\frac{\mathcal{G}_{\rm dm}(z)}{\mathcal{G}_{\rm b}(z)} \frac{\Omega_{\rm dm}}{\Omega_{\rm m}} \right]^2 \Delta^2_{\rm b, PMF}(k,z) \,,
\end{equation}
where $\mathcal{G}_c(z)$ is redshift dependence of growth functions, $\delta^{\rm PMF}_{c}\propto  \mathcal{G}_c(z) S(x)$.}
To make our final results flexible to a more accurate determination of $\kappa$ in the future, we use the above ‘peak’ scale as the PMF parameter which enters our analysis. Another advantage is that different pairs of $\lbrace B_{1\rm Mpc},n_B\rbrace$ with the same $k_{\rm peak}$ value roughly have similar impact on the Lyman-$\alpha$, as we show below.

\begin{figure}
\begin{center}
  \includegraphics[width=0.48\textwidth,trim={0 0 0 0},clip]{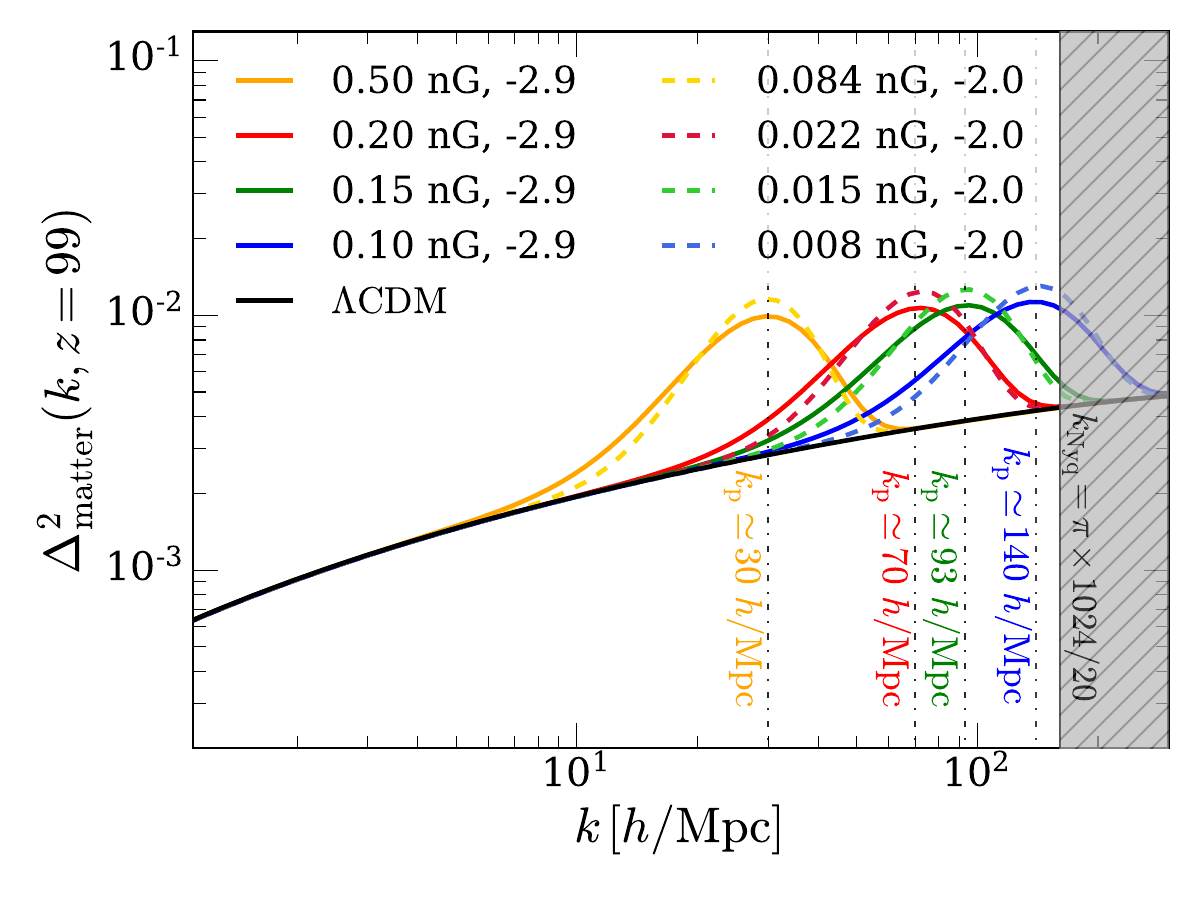}
  \caption{Linear matter power spectra for PMF models with combinations of $\lbrace B_{\rm 1Mpc}, n_{\rm B}\rbrace$ parameters for different peak scales $k_{\rm peak}$. Only spectra with $n_{\rm B} = -2.9$ (full colored lines) are used in all simulations. We also show the corresponding $n_{\rm B}=-2$ spectra (dashed lines) to demonstrate their similarity. Shaded region represents scales beyond Nyquist.} 
  \label{fig:linear_Pk}
\end{center}
\end{figure}

\begin{figure}
\begin{center}
  \includegraphics[width=0.5\textwidth,trim={1.5cm 0.2cm 1.5cm 0},clip]{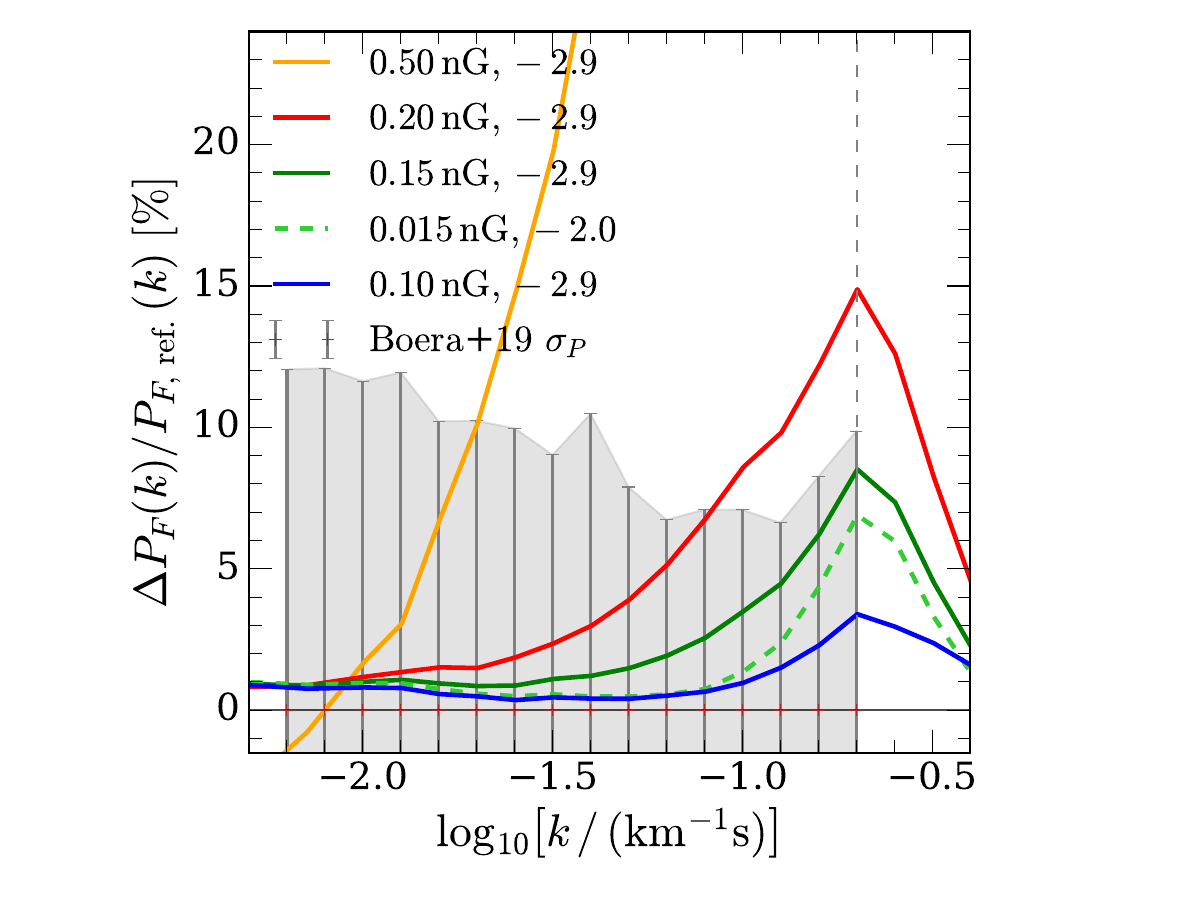}
  \caption{Flux power spectra difference between PMF models at $z=4.6$, where $\Delta P_F = P_F^{\rm PMF} - P_{F,\, {\rm ref.}}$, and  the reference $\Lambda$CDM model. The shaded region with error bars show the observational $1\sigma$ uncertainty.} 
  \label{fig:flux_ratios}
\end{center}
\end{figure}

Here, we consider 4 different PMF models with k$_{\rm peak}=20.1,48,63.3,92.9$ Mpc$^{-1}$ which correspond to $B_{1\rm Mpc}=0.5,0.2,0.15,0.1$ nG (with $n_{\rm B}=-2.9)$. The initial power spectrum is generated at $z=99$, with the same transfer functions for baryons and dark matter (but see \cite{ralegankar24} for another physically motivated approach when using two different transfer functions). In Figure \ref{fig:linear_Pk}, we show the power spectra. To demonstrate potential degeneracy between such models---i.e., whose enhancement of 3D power around $k_{\rm peak}$ could have similar effect in the increase of the 1D Lyman-$\alpha$ flux power---we test two models $\lbrace 0.15, -2.9\rbrace$ and $\lbrace 0.015, -2.0\rbrace$ with $k_{\rm peak} \simeq 93 h/{\rm Mpc}$. It can be seen in the Figure \ref{fig:flux_ratios} that, indeed, their 1D flux power differes by less than 5\%, and is in agreement within the statistical 1$\sigma$ error bars of the data
\\

\textit{Data}--- We use the 1D flux power provided by \cite{boera19}, estimated using 15 high signal-to-noise spectra observed by VLT/UVES \cite{dekker} and Keck/HIRES \cite{vogt}. The measurements span $z= 4.2-5.0$ in bins of $\Delta z = 0.4$. In each bin the flux power spectrum is measured in 15 $k-$bins equidistantly spaced in the range of $\log_{10} k/$ [km/s] $=[-2.2,-0.7]$ with logarithmic spacing of $\Delta \log_{10} k/$[km/s]$=0.1$. The spectrograph resolution is very
high, with $R \sim 50,000$ (FWHM of $\sim 6$ km/s for HIRES
and $\sim 7$ km/s for UVES). The effects of resolution uncertainty are small, even for the highest wavenumber power spectrum bin
measured. Here, we use the measurements with instrumental resolution
corrected, and propagate this correction through the covariance
matrix. The measurements are also corrected for metal contamination. The typical noise estimated in the flux power spectrum is white noise, with an amplitude of $0.1-0.2$ km/s, comparable to the estimated level of the models at the highest wavenumbers. Characterizing the noise levels is important and its impact will be discussed in the Results section.
The $k-$range of the data reaches $k\sim 0.2$ s/km, which would roughly correspond to 50 (physical) kpc at these redshifts. Thus, the data span both high redshifts and the smallest scales, making this an ideal tool to exploit new small scale physics beyond $\Lambda$CDM, including PMFs.
\\
\textit{Hydrodynamic Simulations.}---Capturing non-linear structure evolution of IGM structures and the physics of the gas which gives rise to Lyman-$\alpha$ absorption, such as Doppler broadening and thermal pressure smoothing, requires simulating several different thermal and reionization histories. We rely on a set of high-resolution cosmological hydrodynamical simulations from the Sherwood-Relics project \cite{puchwein23}, using a modified code P-Gadget3 \cite{g2,g4}. This set is complemented by other 48 PMF simulations for the 4 different initial conditions described above. Basically, each PMF model is simulated for 12 different thermal and reionization histories which bracket observational constraints \cite{irsic24}.
The (linear) box size is chosen to be $20\, h^{-1}$Mpc with $2\times 1024^3$ dark matter and gas particles.  We further correct for numerical convergence with both box size and resolution with a series of additional simulations described more extensively in \cite{bolton2017,irsic24} and use a simple and computationally efficient star formation scheme \cite{viel04}.
For the CDM models, we assume a flat cosmology with $\Omega_{\Lambda} = 0.692$, $\Omega_{\rm m} = 0.308$, $\Omega_{\rm b} = 0.0482$, $h = 0.678$, and a primordial helium mass abundance of 0.24. The CDM grid will consist of 108 models which vary thermal history, amplitude and slope of linear matter power around the reference values of $\sigma_8 = 0.829$, $n_{\rm s} = 0.961$, and PMF strength that defines the position of the peak on small scales of the linear matter power spectrum (Fig.~\ref{fig:linear_Pk}).  For each of the simulations we probe the physical parameter space by generating in post-processing a set of $15\times 10 \times 10$ models spanning mean flux values, temperature at the mean density, power-law index of the gas temperature-density relation. The thermal broadening is parameterized by $T_0(z)$ (K) the IGM temperature at mean density  and $\gamma(z)$ (the power-law parameter describing the slope of the temperature-density relation) as  independent parameters at each redshift. As the gas is heated during reionization, it hydrodynamically responds to the resulting increase in its temperature and pressure by expanding. The more heat is injected, the more the gas expands, erasing more small-scale structure. In our models we parametrised this effect with the cumulative heat injected per proton, $u_0(z)$ in eV/m$_{\rm proton}$. 
We also include a template for patchy reionization and quantify the effect of a model dependent resolution correction.
Overall, a suite of 234,000 models for the 1D flux power spectrum for each of the 3 redshift bins probed by the data are fed as an input to our MCMC sampler \cite{data}. 
\
In Figure \ref{fig:flux_ratios} we show the PMF impact on the 1D flux power in terms of percentage difference with the reference $\Lambda$CDM (same thermal history).  The extra power induced by PMFs is very prominent also at the level of 1D flux power and its magnitude and scale dependence can be visually compared with the statistical errors of the data.

\textit{Results.---} We perform a full Monte Carlo Markov Chain (MCMC) likelihood analysis.
In order to validate the robustness of our results we apply several different priors and data cuts. We use the following Gaussian priors on $\sigma_8=0.811\pm0.006$ and on the slope of the linear power spectrum at a pivot scale probed by the data $n_{\rm eff}=-2.3089 \pm 0.005$, and we checked that this does not impact on our results.

The parameter which parameterizes PMFs is defined as $k_{\star}/k_{\rm peak}$ with $k_{\star}=10$ Mpc$^{-1}$, this is done in order to have  the $\Lambda$CDM limit  at zero value, when $k_{\rm peak} \rightarrow \infty$ (similarly to what is done in warm dark matter analyses). We set a conservative flat prior on this parameter in the range [0,0.6] which allows for only a very small extrapolation from our PMF suite of simulations. For this choice the \peak value in Mpc$^{-1}$ would roughly correspond to the magnitude of the B field in nG.

A key prior is the one on the thermal history of the IGM. In this case, we have two different priors: the fiducial (ref) analysis assumes physical priors on the thermal history in the $u_0-T_0$ plane as an envelope around our fiducial grid of simulations; the $u_0-T_0$ priors case refers instead to two independent set of priors on T$_0$ and u$_0$. The difference is that in this latter case the thermal priors have been informed by the measured T$_0$ evolution with redshift from other observational studies, rather than by our suite of hydrodynamic simulations.

\begin{figure}
\begin{center}
  \includegraphics[width=0.45\textwidth,trim={0.2cm 0.2cm 0.2cm 0.2cm},clip]{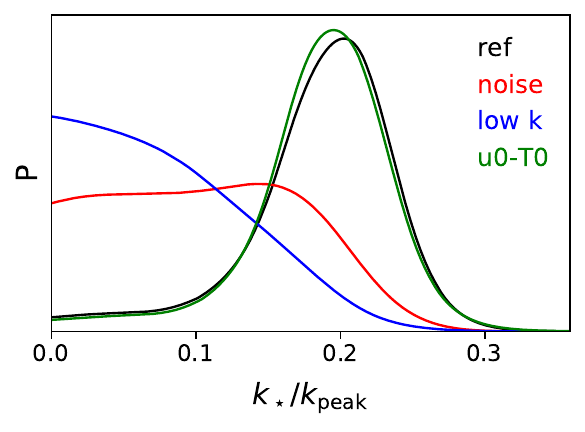}
  \caption{Marginalized 1D distribution for \peak for 4 different PMF analyses: reference case (black), with a physical prior on thermal histories; a conservative treatment of the noise (red); without the small scale data (blue) and with separate priors on temperature and pressure smoothing (green).}
  \label{posterior1D}
\end{center}
\end{figure}

\begin{figure}
\begin{center}
  \includegraphics[width=0.48\textwidth,trim={0.2cm 0.2cm 1cm 1cm},clip]{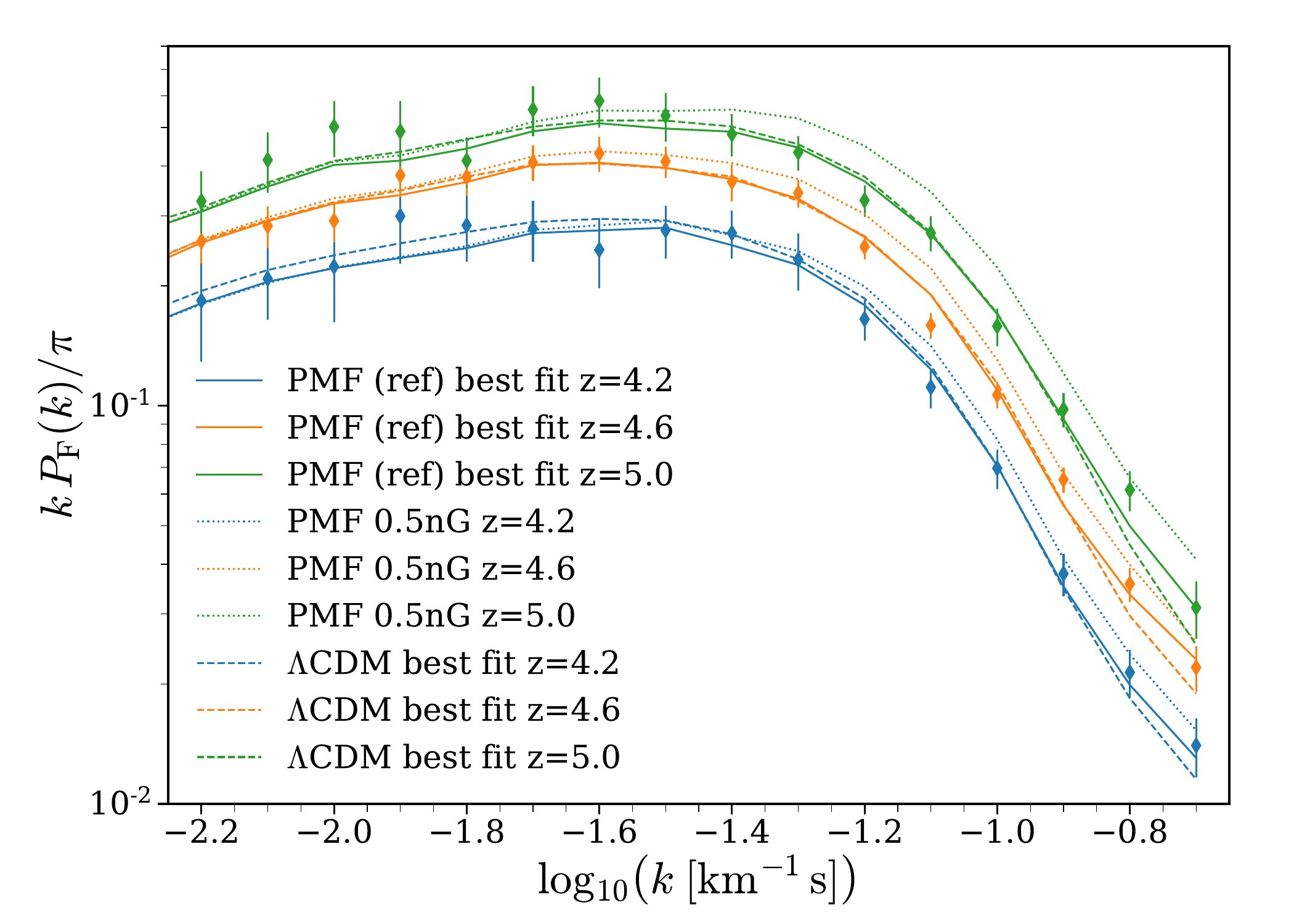}
  \caption{Bestfit PMF and $\Lambda$CDM models, a PMF model with $B=0.5$ nG is also shown (all the other parameters being fixed at the best fit of the PMF ref. case).}
  \label{bestfitPMF}
\end{center}
\end{figure}

Figure \ref{posterior1D} shows the marginalized distribution \peak for the reference case, alternative thermal priors, varying noise, or removal of the four largest value $k-$bins, leaving only data points at $k<0.1$ s/km.
The reference case returns a tentative detection of a peak $k_{\rm \star}/k_{\rm peak} = 0.188^{+0.094}_{-0.12}$ (2$\sigma$ C.L.) with a $\chi^2_{\rm PMF} = 28.63$ for 35 d.o.f. ($p-$value 0.76). Although the posterior is somewhat flat in the range $[0,0.1]$, the $\Lambda$CDM limit (with no PMFs) is excluded at $\sim 3\sigma$. 
For comparison, the $\Lambda$CDM case gives 
$\chi^2_{\rm \Lambda CDM}$ = 40.8 for 36 d.o.f. ($p-$value 0.27).
Thus, the PMF analysis displays a significant  $\Delta \chi^2=-12.2$ improvement with only one extra free parameter. 
In Figure \ref{bestfitPMF}, we highlight that the constraints are especially driven by the smallest scales where PMFs clearly provide a better match to the data. We also show, in order to appreciate the strong impact of PMFs on our observable, a $B=0.5$ nG model which is extremely disfavoured by our analysis.
When we include patchy reionization corrections and resolution corrections that depend on the thermal history we do not observe any appreciable shift in the 1D posterior for \peak.

In order to quantify the robustness of our findings, we also run two extra analyses, explicitly targeted to quantify the crucial role played by the power at small scales.

The first analysis consider a more conservative treatment of the noise, where three noise parameters are added to the theoretical model (one for each redshift bin), and the parameters’ priors are assumed to be
given by the approximate log-normal model of the noise distribution, which is measured from the data (see \cite{irsic24} for more details). These parameters describe a level of residual noise present in the data, and their marginalized values is larger than zero. 
In the reference analysis the noise level is instead added to the theory, without extra parameters. 
When noise is modelled in this way, the 'detection' becomes a robust upper limit $k_{\rm \star}/k_{\rm peak} < 0.25$ (99.7\% confidence level).

If we instead remove the smallest scale data, the bound $k_{\rm \star}/k_{\rm peak} < 0.23$ (99.7\% confidence level) is similar.
It is worth stressing that even in this case, the derived limit is extremely competitive, especially when compared with other probes like the CMB upper limites, or complementary to lower limits from blazars. This implies that also the intermediate and large scales of 1D flux power are sensitive to the increase of power in the linear matter power spectrum, making evident the constraining power of the forest. 

In all our analyses, we  do not find any strong degeneracy between \peak and other parameters, apart from relatively weak degeneracies with the parameters describing the noise and with the injected heat values $u_0(z)$, which is more prominent for $z=4.6,5$.

These results demonstrate that an overall very conservative upper bound on the strength of PMFs at the Mpc scale that will bracket all our analyses could be safely placed as $B<0.3$ nG, since this value is excluded at $3\sigma$ also in the ref PMF analysis.

\textit{Conclusions.---} We have performed an analysis of the 1D Lyman-$\alpha$ flux power with a suite of hydrodynamic simulation which incorporate the PMF-induced increase of power in the linear matter power spectrum. The data displays extra power when compared to the standard $\Lambda$CDM analysis without PFMs, which is well fitted by  a PMF model $B=0.2 \pm 0.05$ nG (1$\sigma$) and $n_{\rm B}=-2.9$.  
However, when a very conservative modelling of the noise level is applied or when removing the smallest scales data, this bound becomes an upper limit $B<0.3$ nG (3$\sigma$). These results are stable  also once different priors on thermal histories and cosmological parameters, patchy reionization, and resolution corrections are considered (see \cite{Supplemental}).
These limits improve the bounds obtained from a combination of CMB data with the inclusion of PMF impact on ionization history by a factor three \cite{Planck:2015zrl} and by an order of magnitude the limits obtained from CMB data alone \cite{paoletti19}, while similar CMB bounds are obtained in \cite{jedamzik19}.
However, we note that \cite{chong13,Pandey2013} by using mean flux measurements and semi-analytical modeling of the forest obtain limits which are a factor $\sim 2$ weaker than those obtained here. Additionally, a more accurate estimate of the damping scale as well as of the baryon and dark matter transfer functions could further affect our results \cite{ralegankar24b}. In this work we neglect the additional heating of IGM that could be produced by ambipolar diffusion and/or MHD turbulent decay (see, for example, \cite{ss2005,jko00,kk14,chluba15,bhaumik25}). This extra heating also impacts the ionization history, potentially influencing at a modest level also the \lya forest \cite{katz21}. 
Our measurement would have exciting consequences in terms of high-redshift galaxy formation and galactic feedback: the dark matter halo mass function will be enhanced below the mass of $10^9\;\mathrm{M_\odot}$$/h$, leading to earlier formation of galaxies. This would have a measurable impact on the interpretation of high-redshift galaxy luminosity functions as observed by the JWST.
In order to discriminate between a bound and a detection, it will be extremely important to further characterize the smallest scales of the 1D flux power by collecting more data at high signal-to-noise and quantify the level of metal contamination. This effort is likely to be achieved easily and on a short time-scale by incrementing the number of high-redshift quasars and increasing the signal-to-noise level to reduce the noise power, using spectrographs like ESPRESSO \cite{espresso}. \\

\textit{Acknowledgments.---}%
 MP, VI and MV are supported by the PD51-INFN INDARK grant. VI is partially supported by the Kavli Foundation. 
 The simulations used in this work were performed using the Cambridge Service for Data Driven Discovery (CSD3), part of which is operated by the University of Cambridge Research Computing on behalf of the STFC DiRAC HPC Facility (www.dirac.ac.uk).  The DiRAC component of CSD3 was funded by BEIS capital funding via STFC capital grants ST/P002307/1 and ST/R002452/1 and STFC operations grant ST/R00689X/1.  DiRAC is part of the National e-Infrastructure. Postprocessing of the simulations was partly performed on Ulysses supercomputer at SISSA. This paper is supported by: the Italian Research Center on High Performance Computing, Big Data and Quantum Computing (ICSC), project funded by European Union - NextGenerationEU - and National Recovery and Resilience Plan (NRRP) - Mission 4 Component 2, within the activities of Spoke 3, Astrophysics and Cosmos Observations. Support by ERC Advanced Grant 320596 ‘The Emergence of Structure During the Epoch of Reionization’ is gratefully acknowledged. MGH has been supported by STFC consolidated grants ST/N000927/1 and ST/S000623/1. JSB is supported by STFC consolidated grant ST/X000982/1. SMA is supported by a Kavli Institute for Particle Astrophysics and Cosmology (KIPAC) Fellowship, and by the NASA/DLR Stratospheric Observatory for Infrared Astronomy (SOFIA) under the 08\_0012 Program. MV is also supported by the INAF Theory Grant "Cosmological Investigation of the Cosmic Web".


\bibliographystyle{bibi}
\bibliography{biblio}

\onecolumngrid
\appendix

\clearpage

\setcounter{equation}{0}
\setcounter{figure}{0}
\setcounter{table}{0}
\setcounter{page}{1}
\makeatletter
\renewcommand{\theequation}{S\arabic{equation}}
\renewcommand{\thefigure}{S\arabic{figure}}
\renewcommand{\thepage}{S\arabic{page}}

\end{document}